\magnification=\magstep1
\baselineskip=0.7056cm
\hsize=16.4truecm \vsize=22.5truecm

\centerline{\bf  Parametric Representation for the  Multisoliton Solution }\par
\centerline{\bf  of the Camassa-Holm Equation}\par
\bigskip
\centerline{Yoshimasa Matsuno
\footnote{$^{a)}$}{{\it E-mail address}: matsuno@yamaguchi-u.ac.jp}} \par 
\centerline{\it Department of Applied Science, Faculty of Engineering,}
\par
\centerline{\it Yamaguchi University, Ube 755-8611} \par
\bigskip
\centerline{March 31, 2005}\par
\bigskip
The parametric representation is given to the multisoliton solution of 
the Camassa-Holm equation. It has a simple structure  expressed in terms of  determinants.
 The proof of the solution is carried out
by an elementary theory of determinants. The large time asymptotic of the
solution  is derived with the formula for the phase 
shift. The latter reveals a new feature when 
compared with the one for the typical
soliton solutions. The peakon limit of the phase shift is also considered,
showing that it reproduces the known result.
  \par
\vfill \eject

\leftline{\bf \S 1. Introduction}\par
In this paper, we report some new results associated with the multisoliton solution of
the  Camassa-Holm (CH) equation$^{1)}$
$$u_t+2\kappa^2u_x-u_{txx}+3uu_x=2u_xu_{xx}+uu_{xxx}.  \eqno(1)$$
Here,  $u=u(x,t)$, $\kappa$ is a positive parameter 
and the subscripts $t$ and $x$ appended to $u$
denote partial differentiation. Originally, this equation has been found in a purely mathematical
search of recursion operators  connected with the integrable  partial differential
equations.$^{2)}$ Recently, 
equation (1)  has attracted  considerable interest since it  has been derived as a model
equation for shallow-water waves.$^{1)}$  In addition,  the  equation has been 
shown to be  completely integrable.  With this as a turning point,
 a large number of works have been devoted 
to studying the mathematical structure of the equation. 
A recent paper describes a short history  and relevant references concerning the CH
and related equation.$^{3)}$  Almost all the works have been focuced on the case $\kappa=0$
for which the CH equation exhibits peakon solutions which are represented by
piecewise analytic functions
and whose dynamics are now well understood.$^{4)}$  However, when $\kappa\not=0$, several new features appear in  solutions. In particular, solutions recover analytic nature, but they are
expressed  by a parametric form  like $u=u(y,t), x=x(y,t)$ 
where $y$ is a new coordinate variable.  A  difficult technical problem is to find the
inverse mapping $x=x(y,t)$. To date, this problem is resolved  only for 
particular cases.$^{3,5,6)}$  In this respect, we remark that an approach based on the
inverse scattering transform method (IST) provides  an explicit form of the inverse mapping 
in terms of Wronskian determinants.$^{7)}$
 Nevertheless, the general $N$-soliton formula is not available yet. \par
The main purpose in this paper is to present  a  complete description  of 
the general $N$-soliton solution  in the form of  parametric representation. We show that 
the solution exhibits a simple structure  expressed in terms of two  determinants. 
In \S 2, we shortly summarize the derivation of a system of equation (associated CH equation)
equivalent to equation (1). In \S 3, we describe the main result in which a parametric
representation is  given to the $N$-soliton solution of the CH equation. The proof
of the solution is carried out by an elementary theory of determinants.
In \S 4, we explore the asymptotic form of the $N$-soliton solution for large
time and show that the interaction process of solitons reveals a new feature
when compared with the typical one for the Korteweg-de Vries equation.
In \S 5, we summarize the results and discuss some open problems related to
the CH equation. \par
\bigskip
\leftline{\bf \S 2. Associated CH equation}\par
It has been demonstrated  that the CH equation can be recast into a more tractable form
by means of  an appropriate coordinate transformation.$^{6)}$ 
Here, we give a short summary of the results.
  Introducing the new
variable $r$  in accordance with the relation
$$r^2=u-u_{xx}+\kappa^2, \eqno(2)$$
the CH equation (1) can be put into the form
$$r_t+(ur)_x=0, \eqno(3)$$
where the boundary condition for $r$ is $r(\pm\infty,t)=\kappa$.
Then, we define the coordinate transformation $(x,t)\rightarrow (y, t^\prime)$ by
$$dy=rdx-urdt, dt^\prime=dt.\eqno(4)$$
In the following analysis, we use the time variable $t$ in place of $t^\prime$ by virtue of
the second relation in (4).  Transforming (3) by means of  (4), it becomes
$$r_t+r^2u_y=0, \eqno(5)$$
and $u$ is expressed in terms of $r$ as
$$u=r^2-r({\rm ln}\ r)_{ty}-\kappa^2.\eqno(6)$$
We term the system of equtions (5) and (6) the associated CH equation.$^{6)}$
If we substitute (6) into (5), we obtain an alternative but more convenient form in the
following analysis, which is 
$$Q_t=r_y,\eqno(7)$$
where
$$Q={1\over 2}{r_{yy}\over r}-{1\over 4}\left(r_y\over r\right)^2
+{1\over 4}\left({1\over r^2}-{1\over\kappa^2}\right).\eqno(8)$$
 By eliminating the variable $r$ from (7) and (8), we can see that $Q$ evolves according
to the following nonlinear wave equation$^{3)}$
$$Q_t+2\kappa^3Q_y+4\kappa^2QQ_t+2\kappa^2Q_y\partial_y^{-1}Q_t
-\kappa^2Q_{tyy} =0,\eqno(9)$$
where $ \partial_y^{-1}=-\int^\infty_ydy$ is an integral operator. An important observation is
that equation (9) can be  identified with a model equation for shallow-water waves.$^{3)}$
 This fact enables us to obtain the $N$-soliton solution of  (9) in the $(y,t)$ coordinate
system. To complete the solution, however, one must revert to 
the original $(x,t)$ coordinate system
via  the inverse mapping
$$x_y={1\over r(y,t)}, x_t=u(y,t). \eqno(10)$$
The most difficult ingredient in the analysis is to integrate (10) for the $N$-soliton solution. \par
\bigskip
\leftline{\bf \S 3. Parametric representation of the $N$-soliton solution}\par
Now,  the main result in this paper is summarized as follows:  The $N$-soliton solution of the CH equation (1) can be written in a parmetric representation
$$  u(y,t)=\left({\rm ln}\ {f_2\over f_1}\right)_t,\eqno(11)$$
$$x={y\over \kappa}+{\rm ln}\ {f_2\over f_1} +d. \eqno(12)$$
Here, $f_1=f_1(y,t)$ and $f_2=f_2(y,t)$ have the determinantal expressions
$$f_1=|G|, f_2=|H|,\eqno(13)$$
with the $N\times N$ marices $G$ and $H$ whose elements are given respectively by
$$G=(g_{ij}), g_{ij}={p_i/q_i\over p_i-q_i}{\rm e}^{\xi_i}\delta_{ij}
+{1\over p_i-q_j}, (i, j=1, 2, ..., N),\eqno(14)$$
 $$H=(h_{ij}), h_{ij}= {q_i/p_i\over p_i-q_i}{\rm e}^{\xi_i}\delta_{ij}
+{1\over p_i-q_j}, (i, j=1, 2, ..., N),\eqno(15)$$
 where
$$\xi_i=(p_i-q_i)y+{\kappa\over 2}\left({1\over p_i}-{1\over q_i}\right)t+\xi_{i0}, 
(i=1, 2, ..., N),\eqno(16)$$
$$p_i={1\over 2\kappa}(1+\kappa k_i),  q_i={1\over 2\kappa}(1-\kappa k_i),
 (i=1, 2, ..., N),\eqno(17)$$
  $\delta_{ij}$ is Kronecker's delta and the parameter $d$ in (12) is an integration
constant.  By virtue of the parametrization (17),
the $N$-soliton solution is characterized completely by 
the $2N$ parameters $k_i$ and $\xi_{i0},\ (i=1, 2, ..., N)$ 
.  In terms of the parameters $k_i$, the phase variable $\xi_i$ of the $i$th soliton may
be written in the form
$$\xi_i=k_i\left(y-{2\kappa^3\over 1-\kappa^2k_i^2}t-y_{i0}\right), (i=1, 2, ..., N),
\eqno(18)$$
 where we have put $\xi_{i0}=-k_iy_{i0}$.   \par
Let us now outline the proof of (11) and (12) in which two bilinear
identities (23) and (38) below will play an essential role.  First, we write the $N$-soliton solution
of equation (9) in a determinantal form$^{3,8,9)}$
$$Q=-2({\rm ln} f)_{yy},  f=|F|,\eqno(19)$$
where $F$ is an $N\times N$ matrix with elements
$$F=(f_{ij}), f_{ij}={{\rm e}^{\xi_i} \over p_i-q_i}\delta_{ij}
+{1\over p_i-q_j}, (i, j=1, 2, ..., N).\eqno(20)$$
Substituting (19) into (7) and integrating the resultant equation by $y$ under the
boundary condition $r\rightarrow\kappa, |y|\rightarrow\infty$, we obtain
$$r=\kappa-2({\rm ln}\ f)_{ty}.\eqno(21)$$
A crucial observation in the present analysis is that $r$ can be represented in terms of $f, f_1$ and $f_2$ as
$$r=\kappa {f_1f_2\over f^2}.\eqno(22)$$
It follows from (21) and (22) that 
$$\kappa f_1f_2=\kappa f^2-2(ff_{ty}-f_tf_y). \eqno(23)$$ 
This is a bilinear identity among determinants.  We note that  analogous relations 
have been studied in the direct
proof of the multiperiodic solutions of the  Benjamin-Ono and 
nonlocal nonlinear Schr\"odinger equations while employing an
elementary theory of determinants.$^{10)}$  
  For later convenience, we first introduce some notations  as well as formulas for determinants and then describe the main result.
Matrices and cofactors associated with any $N\times N$
matrix $A=(a_{ij})$ are defined  as follows :
$$A(a_i; b_i)= \left(\matrix{a_{11}&\ldots&a_{1N}&b_1 \cr
              \vdots&\ddots&\vdots&\vdots \cr
              a_{N1}&\ldots &a_{NN}&b_N \cr
              a_1&\ldots &a_N&0\cr}\right),     \eqno(24)$$
$$A(a_i, b_i; c_i, d_i)= \left(\matrix{a_{11}&\ldots&a_{1N}&c_1&d_1 \cr
              \vdots&\ddots&\vdots&\vdots&\vdots \cr
              a_{N1}&\ldots&a_{NN}&c_N&d_N \cr
              a_1&\ldots&a_N&0&0\cr
               b_1&\ldots &b_N&0&0\cr}\right),    \eqno(25)$$
$$A_{ij}={\partial |A|\over \partial a_{ij}}, 
A_{ij,kl}={\partial^2|A|\over\partial a_{ik}\partial a_{jl}}. \eqno(26)$$
Here, $A_{ij}$ and  $A_{ij,kl}$ are the first and second cofactors, respectively. The following
formulas are used frequently in the present analysis:
 $$\left|\matrix{a_{11}&\ldots&a_{1N}&x_1\cr
               \vdots&\ddots&\vdots&\vdots\cr
               a_{N1}&\ldots&a_{NN}&x_N\cr
               y_1&\ldots&y_N&z\cr}\right|
=|A|z-\sum^N_{i,j=1}A_{ij}x_iy_j,   \eqno(27)$$
$$|A(a_i, b_i; c_i, d_i)||A|=|A(a_i; c_i)||A(b_i; d_i)|-|A(a_i; d_i)||A(b_i; c_i)|, \eqno(28)$$
$$\sum_{i,j=1}^N(f_i+g_j)a_{ij}A_{ij}=\sum_{i=1}^N(f_i+g_i)|A|,\eqno(29)$$
$$\sum_{r,s=1}^N(f_r+g_s)a_{rs}A_{is}A_{rj}=(f_i+g_j)A_{ij}|A|. \eqno(30)$$
Formula (28) is Jacobi's identity and
 formulas (29) and (30) follow from the expansion formulas for determinants,
$\sum_{k=1}^Na_{ik}A_{jk}=\delta_{ij}|A|, \sum_{k=1}^Na_{ki}A_{kj}=\delta_{ij}|A|$.  
\par
Let us now prove (23).   Using the rule for the differentiation of  determinant  and formula
(29) with $A=F$, $f_i=p_i$ and $g_i=-q_i$, we obtain
$$|F|_y=-\sum_{i,j=1}^NF_{ij}+\sum_{i=1}^N(p_i-q_i)|F|. \eqno(31)$$
Similarly, we find from (29) with  $A=F$, $f_i=p_i^{-1}$ and $g_i=-q_i^{-1}$  that
$$|F|_t={\kappa\over 2}\sum_{i,j=1}^N{F_{ij}\over p_iq_j}
+{\kappa\over 2}\sum_{i=1}^N\left({1\over p_i}-{1\over q_i}\right)|F|. \eqno(32)$$
If we use formula (30) with $A=F$, $f_r=p_r$ and $g_s=-q_s$ 
as well as  Jacobi's
identity of the form $|F|F_{li,lj}=F_{ll}F_{ij}-F_{lj}F_{il}$, we can derive the formula
for the $y$ derivative of $F_{ij}$
$$F_{ij,y}=\left\{-\sum_{l,s=1}^NF_{ls}
+\sum_{l=1}^N(p_l-q_l)|F|\right\}{F_{ij}\over |F|}
+\sum_{l,s=1}^N{F_{is}F_{lj}\over |F|}
-(p_i-q_j)F_{ij}. \eqno(33)$$
Differentiating (32) by $y$ and  inserting (31) and (32), we obtain the expression for
$|F|_{ty}$. With (31) and (32), this result is substituted in  the right-hand side
of (23) to obtain the relation
$$\kappa f^2-2(ff_{ty}-f_tf_y)
=\kappa |F|^2-\kappa \sum_{i,j=1}^N{F_{ij}\over p_i}\sum_{l,k=1}^N{F_{lk}\over q_k}
+\kappa|F|\sum_{i,j=1}^N{p_i-q_j\over p_iq_j}F_{ij}.\eqno(34)$$
Further simplication is possible by applying (27) to (34). This gives
$$\kappa f^2-2(ff_{ty}-f_tf_y)=\kappa(|F|+|F(1; p_i^{-1}|)(|F|-|F(q_i^{-1}; 1)|). \eqno(35)$$
On the other hand, using (27) together with the basic formulas for 
determinants, we can show that
$$f_1=|G|=\prod_{j=1}^N(p_j/q_j)(|F|+|F(1; p_i^{-1})|), \eqno(36)$$
$$f_2=|H|=\prod_{j=1}^N(q_j/p_j)(|F|-|F(q_i^{-1}; 1)|). \eqno(37)$$
The identity (23) follows immediately from (35), (36) and (37). \par
The identity below also plays an important role:
$$f^2-f_1f_2=\kappa(f_1f_{2,y}-f_{1,y}f_2), \eqno(38)$$
which we shall now show.
   First, it follows from (36) and (37) that
$$f^2-f_1f_2=|F|(|F(q_i^{-1}; 1)|-|F(1; p_i^{-1})|)
+|F(1; p_i^{-1})| |F(q_i^{-1}; 1)|. \eqno(39)$$
Formulas similar to (31) now take the form
$$|G|_y= \prod_{j=1}^N(p_j/q_j)|F(q_i; p_i^{-1})|+\sum_{i=1}^N(p_i-q_i)|G|, \eqno(40)$$
$$ |H|_y= \prod_{j=1}^N(q_j/p_j)|F(q_i^{-1}; p_i)|+\sum_{i=1}^N(p_i-q_i)|H|. \eqno(41)$$
 Then, we calculate the quantity $P\equiv f^2-f_1f_2+\kappa(f_{1,y}f_2-f_1f_{2,y})$.
Substitution of (36), (37), (40) and (41) into $P$  yields
$$P=|F|(|F(q_i^{-1}; 1)| -\kappa  |F(q_i^{-1}; p_i)|- |F(1; p_i^{-1})|
+\kappa |F(q_i; p_i^{-1})|)$$
$$ +|F(1; p_i^{-1})||F(q_i^{-1}; 1)|
-\kappa |F(q_i; p_i^{-1})| |F(q_i^{-1}; 1)|-\kappa |F(q_i^{-1}; p_i)| |F(1; p_i^{-1})|. \eqno(42)$$
 Let $P_1$ be the sum of terms involving $|F|$ and $P_2$ be the  rest.  Owing to the basic formula
for deteminant, $\alpha|F(a_i; b_i)|+\beta|F(a_i; c_i)|=|F(a_i; \alpha b_i+\beta c_i)|$ 
and the relation $1-\kappa p_i=\kappa q_i$ which follows directly from (17), $P_1$ reduces to
$$P_1=\kappa|F|(|F(q_i^{-1}; q_i)|-|F(p_i^{-1}; p_i)| ). \eqno(43)$$
Using Jacobi's identity and (17), we can show that
$$P_2=\kappa |F||F(p_i, 1; p_i^{-1}, q_i^{-1})|. \eqno(44)$$
After a few manipulations, we finally arrive at the relation
$$P=P_1+P_2={\kappa|F|\over\prod_{i=1}^Np_iq_i}(|\tilde F(q_i; p_i)|- |\tilde F(p_i; q_i)| ),
\eqno(45)$$
where $\tilde F=(\tilde f_{ij})$ is an $N\times N$ matrix with elements
    $$\tilde f_{ij}=p_i(q_if_{ij}+1), (i, j=1, 2, ..., N). \eqno(46)$$
Thanks to (17) and (20),  we  see  that $\tilde F$ is a symmetric matrix and hence
 $|\tilde F(q_i; p_i)|= |\tilde F(p_i; q_i)| $, implying that $P=0$. Thus, we complete
the proof of (38). \par
The relations (22) and (38) immediately lead to our main result. 
Indeed, it follows from (22) and
(38) that
$${1\over r}-{1\over\kappa}=\left({\rm ln}\ {f_2\over f_1}\right)_y.\eqno(47)$$
We rewrite equation (5) in the form
$$ \left({1\over r}-{1\over\kappa}\right)_t=u_y, \eqno(48)$$
and substitute (47) into (48). Integrating the resultant equation in $y$ under the boundary
condition $u\rightarrow 0, |y|\rightarrow\infty$, we obtain the expression (11) 
for $u(y,t)$.  In view of (47),  a system of equations (10) which determine 
the inverse mapping can be
integrated.  This gives rise  to (12).  Note that  the parameter $d$ is 
independent of $t$ as confirmed by (48)
and the second equation in (10). The expressions (11) and (12) give a complete
description of the parametric representation for the $N$-soliton solution of the CH equation.
In the process of constructing soliton solutions, we have noticed that 
it is better to use the determinants $f_1$ and $f_2$  in place of  the determinant $f$.
 It should be pointed out, however, that $f_1$ and $f_2$ exhibit the same functional form as $f$
except phase factors.  Te see this, let $f=f(\xi_1, ..., \xi_N)$. Then,  it follows from (14), 
(15) and (20) that $f_1=f(\xi_1+{\rm ln}(p_1/q_1), ..., \xi_N+{\rm ln}(p_N/q_N))$
and  $f_2=f(\xi_1+{\rm ln}(q_1/p_1), ..., \xi_N+{\rm ln}(q_N/p_N))$.
\par
\bigskip
\leftline{\bf \S 4. Asymptotic behavior of the $N$-soliton solution}\par
A new feature of the $N$-soliton solution appears in the interaction process of solitons.  
Here, we 
address   on this problem.  The procedure for investigating the asymptotic behavior of
the solution can now  be performed straightforwardly. 
 The core part of the calculation is to evaluate $f_1$  and $f_2$
by utilizing the formula for the Cauchy determinant
$$d(m, m+1, ..., n)\equiv\left|\left({p_i-q_i\over p_i-q_j}\right)_{m\le i,j\le n}\right|
=\prod_{m\le i<j\le n}{(p_i-p_j)(q_i-q_j)\over (p_i-q_j)(q_i-p_j)}. \eqno(49)$$
 To this end, we order the magnitude of
the velocity of each soliton in the $(x, t)$ coordinate system  as
$c_1>c_2>...>c_N$ where
$$ c_i={2\kappa^2\over 1-\kappa^2k_i^2}, (i= 1, 2, ..., N). \eqno(50)$$ 
We take the limit $t\rightarrow -\infty$ with the phase variable $\xi_i$ of the $i$th soliton 
being fixed. 
  Since then other phase variables behave like $\xi_1, \xi_2, ..., \xi_{i-1}\rightarrow +\infty,
\xi_{i+1}, \xi_{i+2}, ..., \xi_N\rightarrow -\infty$, 
$f_1$ has the leading-order asymptotic of the form
$$f_1\sim {\rm exp}\left[\sum_{j=1}^{i-1}\xi_i\right]\left(\prod_{l=1}^{i-1}{p_l\over q_l}\right)
\left(\prod_{s=1}^{N}{1\over  p_s-q_s}\right)\times $$
$$\left({p_i\over q_i}{\rm e}^{\xi_i}d(i+1, i+2, ..., N)+d(i, i+1, ..., N)\right).\eqno(51)$$
By invoking  (49) and (17),  we obtain
$${d(i+1, i+2, ..., N)\over d(i, i+1, ..., N)}
=\prod_{j=i+1}^N\left({k_i+k_k\over k_i-k_j}\right)^2. \eqno(52)$$
Substitution of  (52) into (51) gives
$$f_1\sim {\rm exp}\left[\sum_{j=1}^{i-1}\xi_j\right]\left(\prod_{l=1}^{i-1}{p_l\over q_l}\right)
\left(\prod_{s=1}^{N}{1\over  p_s-q_s}\right)d(i, i+1, ..., N)
\left({p_i\over q_i}{\rm e}^{\xi_i-\gamma_i^{(-)}}+1\right), \eqno(53)$$
where
$$\gamma_i^{(-)}=\sum_{j=i+1}^N{\rm ln}\left({k_i-k_j\over k_i+k_j}\right)^2,
(i= 1, 2, ..., N). \eqno(54)$$
Similarly, in the limit of $t\rightarrow -\infty$, $f_2$ has the asymptotic form
  $$f_2\sim  {\rm exp}\left[\sum_{j=1}^{i-1}\xi_j\right]\left(\prod_{l=1}^{i-1}{q_l\over p_l}\right)
\left(\prod_{s=1}^{N}{1\over  p_s-q_s}\right)d(i, i+1, ..., N)
\left({q_i\over p_i}{\rm e}^{\xi_i-\gamma_i^{(-)}}+1\right). \eqno(55)$$
It turns
out from (11), (53) and (55)  that $u$  is represented by a superposition of $N$ solitons
$$u \sim \sum_{i=1}^Nu_i(\xi_i-\gamma_i^{(-)}), \eqno(56)$$
where $u_i$ is a one-soliton solution given by$^{3,5,6)}$
$$u_i(\xi_i)={2\kappa \tilde c_ik_i^2\over 1+\kappa^2k_i^2
+(1-\kappa^2k_i^2)\cosh\ \xi_i}, (\tilde c_i=\kappa c_i). \eqno(57)$$
In the same limit, the mapping relation (12) becomes
$$x-c_it-x_{i0}\sim {\xi_i\over \kappa k_i}
+{\rm ln}\ {(1-\kappa k_i){\rm e}^{\xi_i-\gamma_i^{(-)}}+1+\kappa k_i 
\over (1+\kappa k_i){\rm e}^{\xi_i-\gamma_i^{(-)}}+1-\kappa k_i }$$
$$-\sum_{j=1}^{i-1}{\rm ln}\left({1+\kappa k_j\over 1-\kappa k_j}\right)^2 
-{\rm ln}\left({1+\kappa k_i\over 1-\kappa k_i}\right)+d,
 (i= 1, 2, .., N), \eqno(58)$$
where $x_{i0}=y_{i0}/\kappa$.  In the limit of $t\rightarrow +\infty$,  on the other hand, 
the expressions
corresponding to (56), (54) and (58) are given respectively by
$$u \sim \sum_{i=1}^Nu_i(\xi_i-\gamma_i^{(+)}), \eqno(59)$$
$$\gamma_i^{(+)}=\sum_{j=1}^{i-1}{\rm ln}\left({k_i-k_j\over k_i+k_j}\right)^2,
(i= 1, 2, ..., N), \eqno(60)$$
$$x-c_it-x_{i0}\sim {\xi_i\over \kappa k_i}
+{\rm ln}\ {(1-\kappa k_i){\rm e}^{\xi_i-\gamma_i^{(+)}}+1+\kappa k_i 
\over (1+\kappa k_i){\rm e}^{\xi_i-\gamma_i^{(+)}}+1-\kappa k_i }$$
$$-\sum_{j=i+1}^{N}{\rm ln}\left({1+\kappa k_j\over 1-\kappa k_j}\right)^2
-{\rm ln}\left({1+\kappa k_i\over 1-\kappa k_i}\right) +d,
 (i= 1, 2, .., N). \eqno(61)$$
Let $\Delta_i$ be the phase shift of the $i$th soliton in the $(x, t)$ coordinate system. 
This quantity can be evaluated  simply  with an appropriate use of 
(58) and (61). The result is
$$\Delta_i={1\over \kappa k_i}(\gamma_i^{(+)}-\gamma_i^{(-)})
+\sum_{j=1}^{i-1}{\rm ln}\left({1+\kappa k_j\over 1-\kappa k_j}\right)^2 
-\sum_{j=i+1}^{N}{\rm ln}\left({1+\kappa k_j\over 1-\kappa k_j}\right)^2,
(i= 1, 2, .., N). \eqno(62)$$
The phase shift consists of two contributions. The first term on the right-hand side of (62)
comes from $u(y,t)$  and the rest terms from $x(y,t)$. Note that the first term is the
same as the phase shift arising from the interaction of $N$ solitons for
the KdV  equation.  However, due to the mapping (12), additional terms
appear as indicated by  (62).  
The case $N=2$ in (62) recovers a formula already derived by the IST.$^{11)}$ 
\par
It is also an interesting problem to investigate the characteristics of the $N$-soliton solution
in the limit of $\kappa\rightarrow 0$.  Here, we are concerned only with the limiting form 
of the phase shift.  We find that the appropriate limiting procedure can be performed if one puts
$\kappa k_i=1-\epsilon_i$ and takes the limit $\epsilon_i\rightarrow 0$ while
keeping $c_i (i= 1, 2, ..., N)$.  At the same time,  the conditions 
$\epsilon_ic_i=\epsilon_jc_j (i, j=1, 2, ..., N)$  must be  imposed by virtue of the relation
between $c_i$ and $k_i$ (see (50)).  The formula (62) then reduces to
$$\Delta_i=-\sum_{j=1}^{i-1}{\rm ln}\left({c_i\over c_i-c_j}\right)^2 
+\sum_{j=i+1}^{N}{\rm ln}\left({c_i\over c_i-c_j}\right)^2,
(i= 1, 2, .., N). \eqno(63)$$
In the special case of $N=2$, (63) coincides with a formula presented in ref. 4). \par
\bigskip
\leftline{\bf \S 5. Discussion}\par
We have obtained a simple parametric representation for the $N$-soliton
solution of the CH equation. The solution is expressed compactly by the two 
 determinants
$f_1$ and $f_2$. This finding is of essential importance.
If one calculates  the solution $u$ from (16) and (21) in terms of a single variable $f$, then
 resulting expression becomes a complicated fashion 
  and it would not be of practical use in investigating the structure of the
solution. Introduction of the variables $f_1$ and $f_2$ also leads to an
analytical form (12) for the inverse mapping. 
The explicit form of the $N$-soliton solution makes it possible to construct other class of
solutions. For instance, the rational soliton solutions may be obtained from it by taking
 appropriate long wave limits $k_i\rightarrow 0 (i= 1, 2, ..., N)$.$^{11)}$
\par
Recently, the CH equation has been generalized to a two-dimensional
version by applying an asymptotic expansion method to a system of
water-wave equations.$^{12)}$  It is an interesting problem
to investigating its integrability. The method developed in this
paper may be useful in constructing multisoliton solutions, if
they exist. Furthermore, the Degasperis-Procesi (DP) equation
is a current interest in soliton theory.$^{13,14)}$ Although the DP
equation has a form similar to the CH equation, its mathematical
structure is quite different from that of the CH equation.$^{14)}$
Quite recently, we have succeeded in obtaining the multisoliton
solution of the DP equation by means of a reduction procedure for
the multisoliton solution of the Kadomtsev-Petviashvili equation.$^{15)}$
The solution can be written in a parametric form analogous to the
corresponding solution of the CH equation. However, the simple 
expressions like (11) and (12) are not at hand yet for the general
$N$-soliton solution. This problem is
currently being investigated. \par
\bigskip
\leftline{\bf Acknowledgement}\par
This work was partially supported by the Grant-in Aid for
Scientific Research (C) No. 16540196 from the Ministry of Education, Culture,
Sports, Science and Technology. \par

\vfill\eject
{\bf References}\par
\item{1)} R. Camassa and D. Holm: Phys. Rev. Lett. {\bf 71} (1993) 1661.\par
\item{2)} B. Fuchssteiner and A. Fokas: Physica {\bf D4} (1981) 47.\par
\item{3)} A. Parker: Proc. R. Soc. London {\bf A460} (2004) 2929.\par
\item{4)} R. Camassa, D.D. Holm and J.M. Hyman: Adv. Appl. Mech. {\bf 31} (1994) 1.\par
\item{5)} R.S. Johnson: Proc. R. Soc. London {\bf A459} (2004) 1687.\par
\item{6)} J. Shiff: Physica {\bf D121} (1998) 24.\par
\item{7)} Y. Li and J.E. Zhang: Proc. R. Soc. London {\bf A460} (2004) 2617.\par
\item{8)} M.J. Ablowitz, D.J. Kaup, A.C. Newell and H. Segur: Stud. Appl. Math. 
{\bf 53} (1974) 249. \par
\item{9)} R. Hirota and J. Satsuma: J. Phys. Soc. Jpn. {\bf 40} (1976) 611.\par
\item{10)} Y. Matsuno: J. Phys. Soc. Jpn. {\bf 73} (2004) 3285.\par
\item{11)}  M.J. Ablowitz and J. Satsuma: J. Math. Phys. {\bf 19} (1978) 2180.\par
\item{12)} R.S. Johnson: J. Fluid Mech. {\bf 455} (2002) 63.\par
\item{13)} A. Degasperis and M. Procesi: in {\it Symmetry and Perturbation Theory},
ed. A. Degasperis and G. Gaeta (World Scientific, Singapore, 1999) 22.
\item{14)} A. Degasperis, A.N.W. Hone and D.D. Holm: Theor. Math. Phys. {\bf 133} (2002)
 1463.\par
\item{15)} Y. Matsuno, preprint (February, 2005).\par

 \bye